\documentstyle[11pt,epsfig,epsf,a4]{article}

\begin{document}

\vspace{-2.5cm}
\hfill MKPH-T-97-34
\vspace{1.9cm}

\begin{center}
{\Large \bf Hadron Polarizabilities and Form Factors}\\ \vspace{0.2cm}
{\Large Working Group Summary} \\ \vspace{0.2cm}
D.~Drechsel (Convener)$^1$, J.~Becker$^2$,  A.Z.~Dubni\v ckov$\acute{\mbox{a}}$$^3$, 
S.~Dubni\v cka$^4$,
L.~Fil'kov$^5$,
\\ H.-W.~Hammer$^{1,6}$, T.~Hannah$^7$, Th.~Hemmert$^6$, G.~H\"ohler$^8$, 
D.~Hornidge$^9$, 
F.~Klein$^{10}$, E.~Luppi$^{11}$, A.~L'vov$^5$, U.-G.~Mei{\ss}ner$^{12}$,
A.~Metz$^1$, R.~Miskimen$^{13}$, 
V.~Olmos$^1$, M.~Ostrick$^1$, J.~Roche$^{14}$, S.~Scherer$^1$\\
\vspace{0.5cm}
\begin{it}
$^1$ Institut f\"ur Kernphysik, Johannes Gutenberg-Universit\"at, D-55099 Mainz, Germany \\
$^2$Institut f\"ur Physik, Johannes Gutenberg-Universit\"at, D-55099 Mainz, Germany\\
$^3$Department of Theoretical Physics, Comenius University, Bratislava, Slovak Republic\\
$^4$Institute of Physics, Slovak Academy of Sciences, Bratislava, Slovak Republic\\
$^5$P.N. Lebedev Physical Institute, Moscow, 117924, Russia\\
$^6$TRIUMF, Theory Group, 4004 Wesbrook Mall, Vancouver B. C., Canada V6T 2A3\\
$^7$Institute of Physics and Astronomy, Aarhus University, DK-8000 Aarhus C, Denmark\\
$^8$Inst. f. Theoretische Teilchenphysik, Universit\"at Karlsruhe, D-76128 Karlsruhe, Germany\\
$^9$SAL, University of Saskatchewan, 107 North Road, Saskatoon, SK, Canada\\
$^{10}$Physikalisches Institut, Universit\"at Bonn, D-53115 Bonn, Germany\\
$^{11}$Universit$\acute{\mbox{a}}$ and INFN, Ferrara, Italy\\
$^{12}$Institut f\"ur Kernphysik, Forschungszentrum J\"ulich, D-52425 J\"ulich, Germany\\
$^{13}$Dep. of Physics and Astronomy, University of Massachusetts,
Amherst, MA 01003, USA\\
$^{14}$DAPHNIA-SPhN, CE Saclay, France\\

\end{it}
\end{center}

\section{Introduction}

The goal of the group was to summarize
the current status of our knowledge on form factors and polarizabilities of
hadrons, and to identify experimental and theoretical activities which could
give new and significant information. These were 4 days of stimulating talks
and discussions. The agenda was a mixture of experiment and theory with the
aim to have frank and critical discussions on the current issues. The topics
covered by the working group are presented as follows.

The status of hadron
form factors is reviewed in Section 2. Experimental data 
for the nucleon were presented for both time-like and 
space-like momentum transfers by E.~Luppi and F.~Klein. These two experimentally
disjunct regions are connected by dispersion relations as discussed by
G.~H\"ohler, A.~Z.~Dubni\v ckov$\acute{\mbox{a}}$ and S.~Dubni\v cka. U.-G.~Mei{\ss}ner presented new
calculations of the spectral functions in the framework of ChPT, and
H.-W.~Hammer reported on a dispersion calculation for the strangeness vector
current. The relevance of double polarization for improving on the neutron
data basis was stressed by J.~Becker and M.~Ostrick. Finally, T.~Hannah 
reported on calculations of the pion form factor by the inverse amplitude
method.

Section 3 is devoted to the polarizability of hadrons as seen by Compton
scattering. V.~Olmos described a Mainz experiment supposed to yield new 
precision
values for the (scalar) polarizabilities of the proton, and D.~Hornidge
showed new data for Compton scattering off the deuteron. The current status
of ChPT calculations of the spin (or vector) polarizabilities was reviewed
by Th.~Hemmert. Finally, L.~Fil'kov described a Mainz experiment to measure
the pion polarizability by radiative pion photoproduction, 
and A.~L'vov reported
on a proposal to study the Compton amplitude by lepton pair production.

Section 4 covers the field of virtual Compton scattering (VCS), which has the
potential to give information on the spatial distribution of the polarizabilities.
S.~Scherer gave an introduction to a VCS calculation in the framework of ChPT,
and A.~Metz reported on relations between the generalized polarizabilities
(GPs) measured by VCS. A designed experiment to measure the GPs at MIT/Bates
was presented by R.~Miskimen, and J.~Roche described the preliminary results
of such measurements at MAMI.

\section{Hadron Form Factors}
The form factors of hadrons have been studied by electron scattering and pair
annihilation or creation. The virtual photon exchanged in these reactions has
four-momentum q, and $t=q^2$ defines the Mandelstam variable. 
In the case of pair annihilation the momentum transfer is time-like, 
$q^2 \ge 4m^{2}_{N}$,
electron scattering probes the form factors at space-like
momentum transfer, $q^2 < 0$. Fig. 1
shows the existing data and a model-calculation for the magnetic form factor
of the proton for both time- and space-like momentum transfers \cite{Ham96}. 
The form factor
 is real for $t < 4m_{\pi}^{2}$ or $9m_{\pi}^{2}$ for its isovector or 
isoscalar part, respectively,
it becomes complex above these lowest thresholds corresponding to two- or 
three-pion production. The spikes in the unphysical region 
$0 \le t \le 4m_{N}^{2}$
correspond to vector mesons, which dominate the imaginary part of the form
factor.

\begin{figure}[h,t,b]
\begin{center}
\epsfig{file=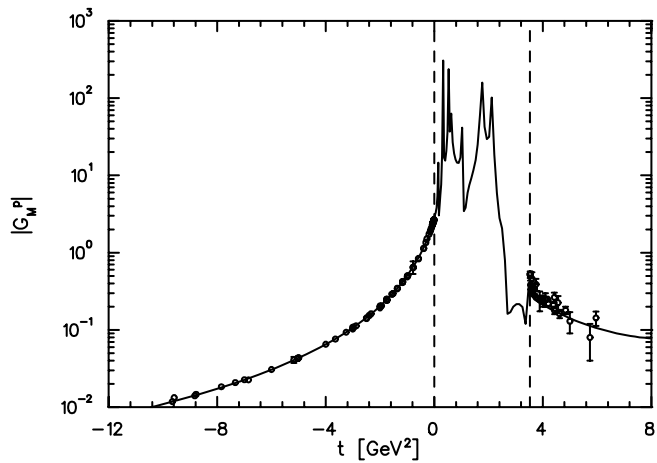,width=6cm, angle=90}
\caption{\small The magnetic form factor of the proton $\mid G_M^p \mid$ 
as function of $t$. See text.}
\end{center}
\end{figure}

The status of the nucleon form factors in the time-like region was reviewed by
E. Luppi. Data were presented from both the reaction 
$N\bar{N} \rightarrow e^{+}e^{-}$ (E760, PS170) \cite{Arm93, Bar94} 
and the inverse reaction (FENICE)\cite{Ant94}. The differential
cross section for this process defines the electric $(G^{N}_{E})$ and magnetic
$(G^{N}_{M})$ Sachs form factors of the nucleon \cite{Zic62},

\begin{equation}
\frac{d\sigma}{d\Omega} = \frac{\alpha^{2}\beta_{N}}{4t} 
 \left( |G^{N}_{M}(t)|^2
  (1+\cos^{2}\theta) + \frac{4m^{2}_{N}}{t}| G^{N}_{E}(t)|^{2} \sin^{2}\theta
 \right),
\end{equation}

with $\beta_{N}$ the velocity of the nucleon and $\theta$ the production
angle. Up to now the angular
distributions have been measured only with large error bars, particularly in
the case of the neutron. Taken at face value, the form factors of proton (p) and
neutron (n) are related by the data as follows:

\begin{equation}
 |G^{p}_{E}| \approx |G^{p}_{M}| ,\  \ |G^{p}_{M}| < |G^{n}_{M}| , 
 \ \ |G^{n}_{M}| \gg |G^{n}_{E}|.
\end{equation}

Since the electric and magnetic Sachs form factors at threshold are equal by
definition,

\begin{equation}
 G^{N}_{E} (4m^{2}_{N}) = G^{N}_{M} (4m^{2}_{N}),
\end{equation}

the relations eq.(2) would require a resonating behaviour of $G^{n}_{M}$
very close to threshold. In general the magnetic form factors are better known,
because the electric ones are suppressed by kinematical factors in 
in most experiments. 
The experimental results for the proton form factor in the time-like region
are presented in Fig. 2. In particular they include the experiments $PS 170$
at CERN \cite{Bar94} in the range between threshold and $q^2 = 4.2 GeV$,
and $E760$ at Fermilab \cite{Arm93} in the region $8.9 \le q^2/GeV^2 \le 13$.
The FENICE results are obtained by $e^+e^-$pair annihilation, and evaluated 
under the hypothesis $\mid G_E^p \mid = \mid G_M^p \mid$.
Though the threshold data have still large
error bars, the indication of a ``resonance'' in the form factors near
threshold is quite strong. This may be related to a dip in the total cross
section for $e^{+}e^{-} \rightarrow$ hadrons at about 1.87 GeV with a small
width of about 10 MeV.
Another point of interest is the predicted asymptotic scaling of the form
factor, i.e. $q^{4}|G| \rightarrow$ const for large values of $q^{2}$. 
As may be seen in Fig. 2 this quantity is rather constant in the region 
$q^{2} > 8GeV^{2}$. However, it stays at about
twice the level reached at the corresponding momentum transfer in the space-like
region (i.e. for $q^{2} < -8 GeV^{2}$)! This is a clear indication that
asymptotia is not yet reached in the region of
$10-20 GeV^{2}$.

\begin{figure}[h,t,b]
\begin{center}
\epsfig{file=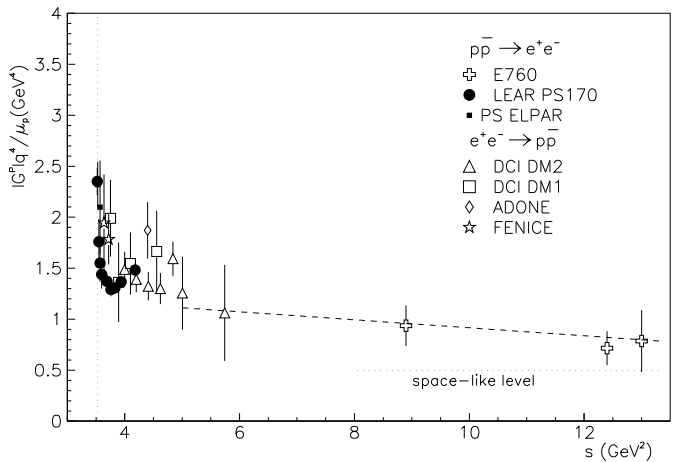, width=9cm}
\caption{\small The magnetic form factor of the proton 
$\mid G^p \mid q^4 / \mu_p$ as function of $s = q^2$
in the time-like region.See text.}
\end{center}
\end{figure}

The form factor of the neutron is shown in Fig. 3. The result at $\sqrt{s} = 1.9 GeV$ is shown as a shaded area corresponding to two different hypotheses
for the $c. m.$ energy of the accelerator. The neutron form factor is considerably larger than the proton one over the whole energy range and, surprisingly,
the magnetic neutron form factor is found to be much larger than the electric
one.

Clearly, the unexpected results in the time-like region deserve
future attention. Therefore, the Fenice Group is studying possibilities to improve
the data by a high luminosity, asymmetric $e^{+}e{-}$ collider with a high
energy $e^{+}$ storage ring and a high intensity $e^{-}$ linac.

\begin{figure}[h,t,b]
\begin{center}
\epsfig{file=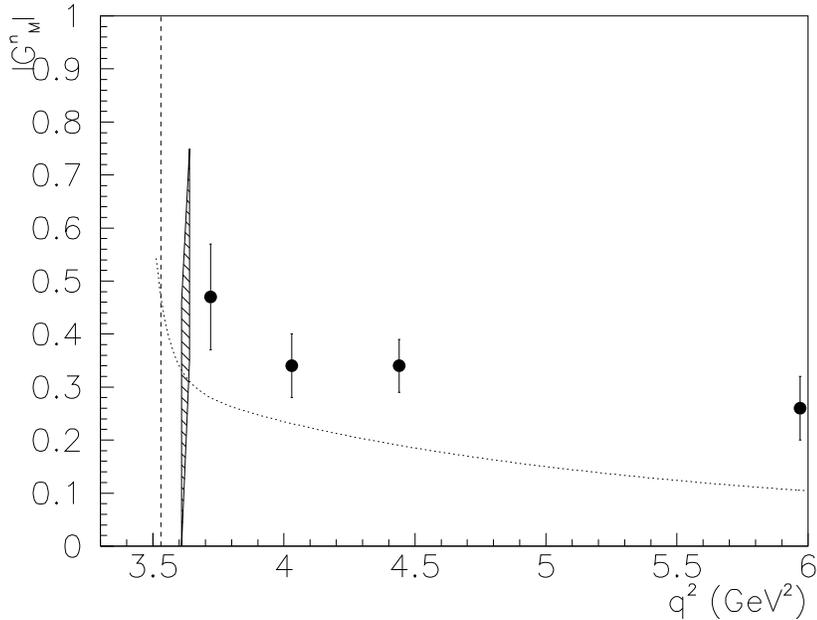, width=11cm}
\caption{\small The magnetic form factor of the neutron $\mid G_M^n \mid$
as function of $q^2$ in the time-like region. See text.}
\end{center}
\end{figure}

F.~Klein reported on the status of the space-like nucleon form factor. The
differential cross section in this region is given by

\begin{equation}
 \frac{d\sigma}{d\Omega} = \sigma_{p} \left (\frac{G^{2}_{E} + \tau G^{2}_{M}}
  {1+\tau} + 2\tau G^{2}_{M} \tan^{2} \frac{\theta}{2} \right ),
\end{equation}

with $\tau = Q^{2}/4m^{2}_{N}$,
$Q^2 = -q^2$, and $\theta$ the scattering angle. Except for the
electric form factor of the neutron, the form factors follow the dipole shape,
  $G^{p}_{E} \approx G^{p}_{M}/\mu_{p} \approx G^{n}_{M}/\mu_{n} \approx
   (1 + Q^{2}/0.71 GeV^{2})^{-2}$.
From a nonrelativistic, and somewhat 
questionable point of view, this form factor corresponds to an exponential 
charge distribution as function of the radius r,
$\rho(r) = \rho_{0}exp (-r/r_{0})$.

Due to the kinematical factors in eq. (4), $G^{p}_{E}$ dominates at small 
values
of $Q^{2}$ and $G^{p}_{M}$ at large ones. This makes it difficult at both small
and large $q^{2}$, to separate the two form factors by studying the angular
dependence at $Q^{2} = const$ (``Rosenbluth plot''). 
For the proton precise electron scattering data are available from single
arm experiments up to momentum transfers $Q^2 = 30 GeV^2/c^2$. A Rosenbluth
separation was made \cite{Bos92, Sil93} up to $9 GeV^2/c^2$ while 
for higher $Q^2\;,\;G_M$ was 
directly taken from the cross section with the small $G^p_E$-contribution
$(\le 5 \%)$ subtracted. This limit may be pushed upwards in $Q^2$ in double
polarization experiments which are planned or in preparation. In particular,
the data at lower $Q^2$ can be substantially improved by means of double
polarization experiments at the new electron accelerators in the GeV region. It
is worth pointing out that even a new measurement of the electric $rms$ radius
of the proton may be interesting. This quantity was determined by a series of 
Mainz experiments in the 70's as $r^{p}_{E} = (0.862 \pm 0.012) fm$ 
\cite{Sim80}. It tends to
be somewhat smaller in a global fit to the data performed in the framework of
dispersion theory, $r^{p}_{E} = (0.847 \pm 0.008) fm$ \cite{Mer96}. However, these small
uncertainties are now the limiting factor in atomic physics investigations to
check the validity of QED \cite{Wei94}!

For obvious reasons the situation is worse in the case of the neutron. Precise
numbers exist only for its electric radius as determined by the transmission of
low energy neutrons through Pb atoms \cite{Kop95},

\begin{equation}
 \left <r^2 \right >^n_E = (-0.113 \pm 0.003 \pm 0.004) fm^2,
\end{equation}

indicating a small value for the Dirac form factor $F_1^n$. On the contrary,
at high momentum transfers the observed smallness of $G_E^n$ \cite{Lun93}
demands a finite value for $F_1^n$.
The older neutron data rely mostly on quasi free scattering off the neutron
bound in $^2H$. In that approximation the longitudinal and transverse structure
functions determine $\left (G^n_E \right )^2$ and $\left (G^n_M \right )^2$,
respectively. Since $G^n_E(0) = 0$ and $G^n_E \ll G^p_E$ at all the measured 
$Q^2$ values, the electric neutron form factor obtained by this method is
essentially zero with a large error bar. Concerning $G^n_M$, the SLAC data 
\cite{Bos92}
cover the region between $1.5 GeV^2$ and $4GeV^2$ with relatively small error
bars. At the smaller momentum transfer the data scatter considerably. Though
there exist more recent data from NIKHEF \cite{Ank94}, Mainz \cite{Jou97} 
(preliminary) and Bonn \cite{Bru95} with
small error bars, the Bonn data are systematically higher by about $10\%$.
It is obvious that this discrepancy requires further investigations.

Relatively precise values for $G^n_E$ were obtained by elastic
scattering off the deuteron under the assumption
that the contributions of the proton,
the internal wave function and two-body currents are well under control 
\cite{Pla90}. In
order to obtain more model-independent results, double polarization 
experiments have been proposed with the aim to measure asymmetries, e.g.

\begin{equation}
 ^3\vec{H}e (\vec{e},e'n),\ \  D(\vec{e},e'\vec{n}) \quad \mbox{and}
 \quad \vec{D}(\vec{e},e'n).
\end{equation}

The first of these reactions has been studied at Mainz and the results were
presented by J.~Becker, the second one has been scheduled at MIT/Bates and
measured at Mainz (preliminary results presented by M.~Ostrick), and the
third one is being pursued at Jefferson Lab.

As pointed out by Ref. \cite{Arn81}, the 
asymmetry in a double polarization experiment takes the form

\begin{equation}
 A = P_eP_n \frac{a G^{n}_{E} G^{n}_{M} \sin \Theta +b (G^n_M)^2 \cos \Theta}
               {c (G^n_E)^2 + d (G^n_M)^2},
\end{equation}

where $P_e$ and $P_n$ are the polarization of the incoming electron and the 
target or recoiling neutron, respectively, and a, b, c, d are known
kinematical functions. The angle $\Theta$ is the angle between the momentum
$\vec{q}$ of the virtual photon and the neutron spin in the scattering plane
of the electron, i.e. $\Theta = 0$ if the neutron spin points in the direction
of $\vec{q}$, and $\Theta = 90^{\circ}$ if perpendicular to $\vec{q}$, in the
scattering
plane and to the same side as the momentum of the outgoing electron. In the 
latter kinematics the asymmetry $A_{\perp}$ essentially measures 
$G^n_E / G^n_M$,
while the parallel asymmetry $A _{\parallel} $ is practically independent of 
the form
factors because of the small value of $G^n_E$. Particularly in the case of the
deuteron target, there are convincing arguments that the analysis of these 
double
polarization experiments is only little affected by two-body currents and
binding effects \cite{Are87}. In this sense the new data for $G^n_E$ 
should be rather model-independent. 

Concerning the reaction $D (\vec{e}, e' \vec{n})$ at MAMI \cite{Ost97},
the scattered electrons were detected in an array of lead glass detectors. 
The energy 
information $(\Delta E/E \approx 20 \%)$ was used to suppress non quasifree
events. The neutron were detected in coincidence in a wall of plastic 
scintillators. The detection process itself provides the analyzing power for the extraction of polarization components transverse to the direction of
neutron momentum. An experimental calibration of the polarimeter at a
polarized neutron beam can be circumvented by using the spin precession in
a magnetic field. On their way through a magnet with a field perpendicular
to the scattering plane, the neutrons undergo a spin precession. The angle
$\Theta_0$ of zero crossing of the asymmetry $A$ of eq. (7) is directly 
related to the ratio of the electric and magnetic form factors,

\begin{equation}
\tan \Theta_0 = - \frac{a}{b} \cdot \frac{G_E}{G_M},
\end{equation}

and depends neither on the absolute values of the polarimeter's analyzing 
power nor on the electron beam polarization. Kinematical cuts change the
amplitude, i.e. the effective analyzing power, but not the zero crossing
angle $\Theta_0$. 

The reaction $^3\vec{He}(\vec e,e'n)$ was measured at MAMI 
in the range of momentum transfer Q$^2$=0.27-0.5 GeV$^2$. 
$^3\vec{He}$ serves as an effective polarized
neutron target with its spin oriented in perpendicular or
parallel direction to the momentum transfer.
The experiment quantifies the respective asymmetries 
whose ratio $A_{\perp}/A_{\|}$ 
is independent of the polarization degrees of electron beam 
and target and of a dilution of the two asymmetries due 
to unpolarized background. The PWIA analysis is based on
an event--by--event kinematical reconstruction that is supported 
by a Monte Carlo simulation. The analysis considers the 
kinematical acceptance of the detector system 
($\Delta \Omega_e$=100 msr) and provides an efficient
selection of quasi--elastic scattering with low missing 
momentum ($|\vec{P}_m|$$\le$100~MeV/c). The final step of
this analysis is a Monte Carlo adaptation 
of the reconstructed kinematics which takes account 
of the effects that can not be determined for the single event 
but are known from their statistical distributions: missing energy, 
radiation losses and energy losses of the neutrons in the lead 
shielding of the neutron detector. Test measurements with 
H$_2$-- and D$_2$--filled target cells examined the 
contribution to the signal due to (p,n)--charge conversion 
in this lead shielding which could be reduced to a fraction
of 4\% in the event--by--event reconstruction. With a 
value of 1.05$\pm$0.05 times the dipole fit for G$_{M}^n$ the 
Q$^2$--averaged result is G$_{E}^n$=0.0352$\pm$0.0033$\pm$0.0024 \cite{Bec97}.
This result is shown in Fig. 4 together with an earlier exploratory $^3$He--experiment 
\cite{Mey94}, the preliminary results of the deuteron experiments \cite{Ost97} 
and data extracted from elastic electron scattering off 
deuterium \cite{Pla90}. The 
four lines in the plot are two-parameter fits to the 
data set of \cite{Pla90} by use of different 
nucleon-nucleon potentials in the calculation 
of the deuterium wave function. Data points are 
plotted for the case of the analysis with the Paris 
potential (fit: solid line). 
The $^3$He experiment will be continued at 
higher \mbox{momentum} transfer, $Q^2$=0.67~GeV$^{2}$, at 
the MAMI spectrometer facility (A1--collaboration) 
\cite{Hei95}.

\begin{figure}[t]
\begin{center}
\epsfig{file=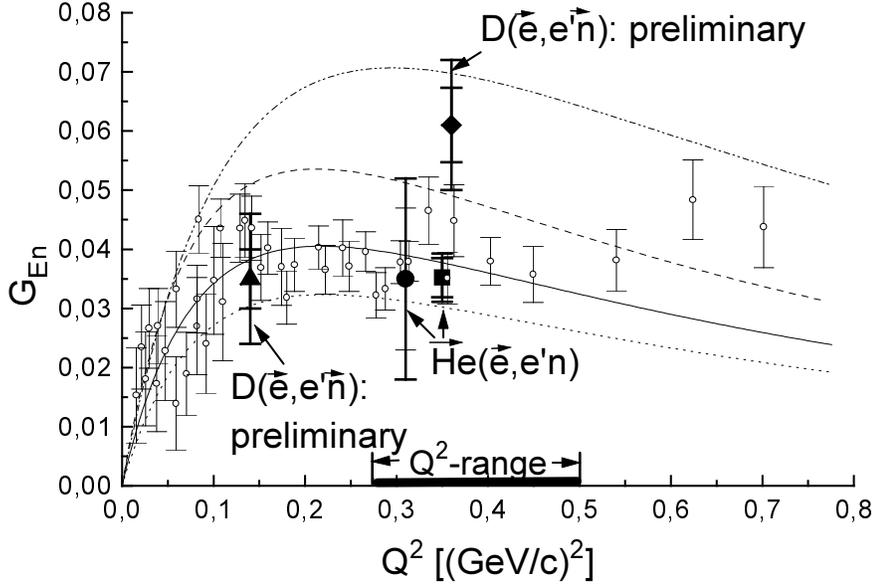,width=8cm,angle=-90}
\caption{\small The electric form factor of the neutron
$G_E^n$ as function of $Q^2$. Full circles: exploratory
$^3$He experiment \cite{Mey94}; full squares: full 
$^3$He experiment (stat. and quadr. added total error) 
\cite{Bec97}; 
full triangle and full diamonds: D experiment \cite{Ost97}, 
preliminary; lines and open points: elastic $D(e,e')$ 
\cite{Pla90}, see text.}
\label{gen}
\end{center}
\end{figure}

As may be seen in Fig. 4, the preliminary results for the
deuteron are higher than for the $^{3}He$ experiment. There are three possible
explanations for this effect, systematical errors in one or the other experiment,
problems with final state interactions and two-body currents in $^{3}He$ (a
complete Faddeev calculation for the reaction is still missing), and possible
modifications of the nucleon's form factors in the medium. Of course, the latter
 possibility would be a most interesting effect which has been under discussion
and looked for in many investigations. With regard to such medium effects one
has to keep in mind that the factor 2 between the two experimental values at
$Q^2 \approx .35 GeV^2$ could be achieved by shifting only a relatively small
amount of charge in the neutron, because the positive core and the negative
cloud are of equal size and small. A similar change of the form factor of the
proton by an absolute value of about 0.03 would be much less exciting. In
conclusion it is obvious that $G^n_E$ is particularly sensitive to the details
of models describing the hadronic structure. Therefore, further activities
to measure this quantity with better accuracy and improved data analysis 
should be strongly encouraged.

The most quantitative description of the electromagnetic form factors can be
obtained within the framework of dispersion theory for the Dirac and Pauli 
form factors
$F_1$ and $F_2$, e.g.

\begin{equation}
 F_{2}(t) = {1\over \pi} \, \int_{t_0}^\infty \,{{\rm Im}~F_{2}(t') \over
 t'-t-i\varepsilon} \,\, ,
\end{equation}

where the spectral function (SF) ${\rm Im}~F(t)$ encodes the pertinent
physics. It is well known that the isovector SF has a significant
enhancement on the left wing of the $\rho$ meson due to a singularity on the
second Riemann sheet at $t_c= 4m_\pi^2 - m_\pi^4/m_N^2$ close to the
threshold $t_0 = 4m_\pi^2$ (for a detailed discussion, see \cite{Hoe83}). 
This is also found in chiral perturbation
theory (CHPT) at one loop as first observed in \cite{Gas88}. One can
therefore ask the
question whether a similar phenomenon appears in the isoscalar
channel, i.e. on the left wing of the $\omega$ meson. For that, one has to
analyze the two--loop diagrams depicted in Fig. 5 \cite{Ber96}. Although the
solution of the pertinent Landau equations reveals a singularity on
the second sheet at $t_c = 8.9 m_\pi^2$, i.e. close to the physical
isoscalar threshold at $t_0 = 9 m_\pi^2$, there is no enhancement as
shown in Fig. 5 (right panel) due to the three--body phase space
factors. The nonresonant part of the SF related to the three--pion
continuum is very small and rises smoothly with increasing $t$.
Consequently, it is justified to saturate the isoscalar SF at
low $t$ by the $\omega$. The same is found for the SF of the
nucleon isovector axial from factor \cite{Ber96}.

\begin{figure}[ht]
\hspace{2.6truein}
\vspace{0.7cm}
\epsfysize=1.8in
\epsffile{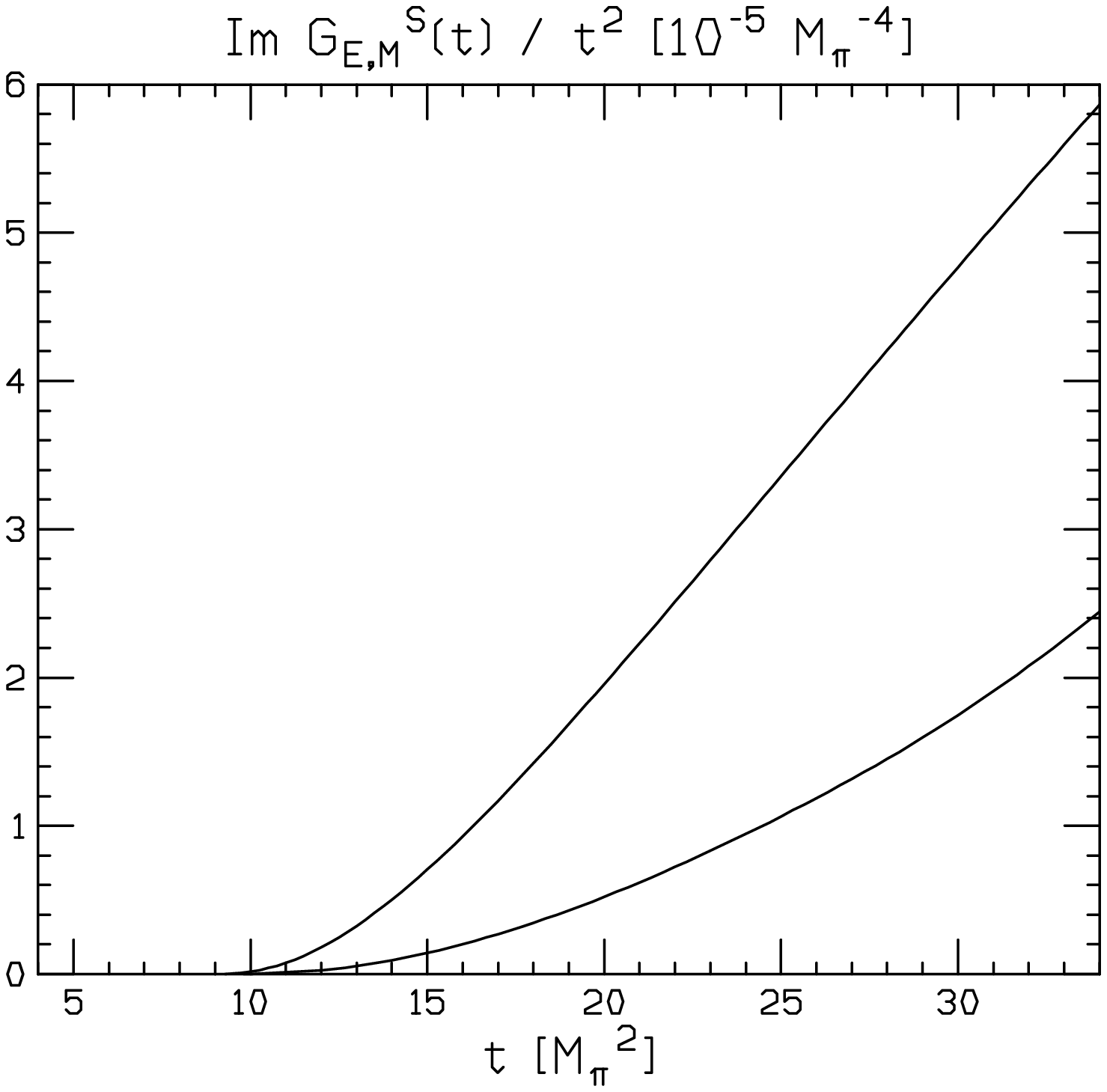}
\vskip -2in
\epsfysize=1.6in
\hskip .1truein
\epsffile{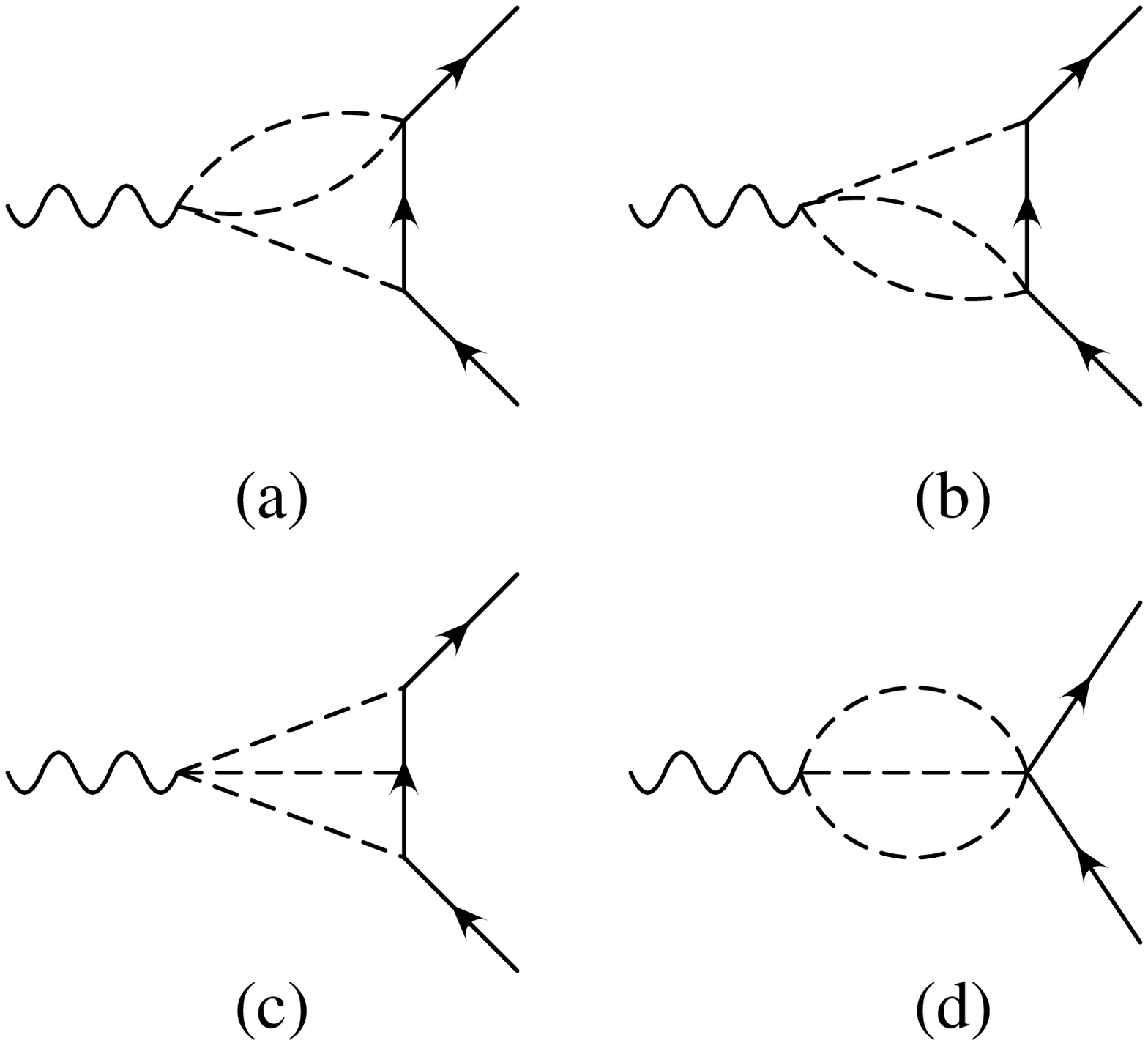}
\vspace{-0.1cm}
 \caption{\small
      Right: Isoscalar spectral functions weigthed with $1/t^2$ for the 
      electric
      (lower curve) and magnetic (upper curve) Sachs form factors. 
      Left: the underlying two--loop graphs (solid,
      dashed, wiggly lines: nucleons, pions and photons). Calculations by
      Bernard, Kaiser and Mei{\ss}ner \cite{Ber96}.
    \label{exfig} }
\end{figure}

The first peak of the spectral function of the isoscalar vector form factor
is due to $\omega$ exchange. Its height is proportional to the coupling
constant $g_1(\omega N N)$. One would expect that the values of this coupling
constant found in analyses of different groups get closer with the recent
improvements of the experimental data base. Unfortunately, this is not true.
The values show large fluctuations even in recent years.

The main reason was pointed out in \cite{Hoe76}: the calculation of the
spectral function from the data is a mathematically {\em ill-posed problem}. If
a fit with a ``good'' $\chi^2$ has been found, there exists a large number of
other solutions with the same $\chi^2$ but quite different spectral functions
which usually show additional oscillations. In Ref. \cite{Hoe76} a {\em stable
analytic continuation} was applied which was justified in Ref. 
\cite{Sab80}. The number of adjustable parameters was reduced by calculating
the $\rho$ contribution from a continuation of $\pi N$ amplitudes and p-wave
$\pi \pi$ phase shifts.

In a recent paper \cite{Mer96} this method was improved and new data were
taken into account. The authors obtained $g_1^2(\omega N N)/4\pi = 34.6\pm
0.8$, close to the old value. Other recent results vary in the range from 4 to
363. Some of the reasons for this were pointed out in Ref. \cite{Hoe93}, and one should
keep in mind that structures of spectral functions are not
necessarily due to vector mesons and that the experimental information on
higher $\rho$ and $\omega$ states is rather poor.

Concerning the values of coupling constants determined from nucleon-nucleon
scattering,  $g_1^2(\omega N N)/4\pi = 8.1\pm 1.5$ was found in a careful
study of the dispersion relations for $N N$ forward scattering \cite{Gre80}. The
$\phi$ contribution was assumed to be zero. Otherwise the $\omega$ coupling
would become {\em smaller}. Results from OBE models of the NN force have the
problem that, in contrast to couplings in dispersion theory, the $\omega$ is
off-shell, at $t\approx 0$ rather than at $t = m^2_V$. The extrapolation to
the on-shell value is very model-dependent, leading to typical values of
$g^2_1/4\pi = 20$ \cite{Mac87}. Obviously the large coupling constant found by
form factor analyses would require a larger $\omega NN$ transition radius than
used in OBE models.

The slope of the Dirac isoscalar form factor at $t=0$ can be obtained rather
accurately from the
slope of the electric Sachs form factor of the proton and the form
factor of the neutron, $F'_{1s}(0) = 1.36\, GeV^{-2}$. The first two terms
of the result of Mergell et al. \cite{Mer96} due to the $\omega$ 
and the $\phi$ terms are $F'_{1s}(0) = (2.00 - 0.683) = 1.32\, GeV^{-2}$. 
If we use the coupling constant of Grein et al., the $\omega$ contribution
is only 1.00 and therefore {\em smaller than the experimental value}. It is
then not possible to describe that $F_{1s}(t)$ behaves approximately like the
usual dipole fit.

A.~Z.~Dubni\v{c}kov$\acute{\mbox{a}}$ explained that the large $\omega$ coupling comes about if
the dipole form is constructed by $\omega$ and $\phi$ poles, while a much 
smaller
coupling is needed if the dipole is made up of $\omega$(782) and $\omega'$
(1420).
Under the usual conditions of normalization and asymptotic behaviour of the 
form
factor, she obtained $g^2_1(\omega NN)/4\pi \approx 34$ for $\omega$ and
$\phi$ poles, and $g^2_1(\omega NN)/4\pi \approx 12$ for $\omega$ and $\omega'$
poles. Of course the latter assumption requires to introduce a strong
coupling to the $\omega'$.

S.~Dubni\v{c}ka presented a model of the form factors based on the resonances
$\omega$(782), $\phi$(1020), $\omega'$(1420), $\omega''$(1600), $\phi'$(1680) 
and
$\rho$(770), $\rho'$(1450), $\rho''$(1700), $\rho''' $(2150), and $\rho''''$
(2600).
All these resonances are experimentally confirmed, except for the
$\rho'''$(2150), taken from \cite{Bia91} and the $\rho''''$(2600) which 
has been
postulated to reproduce the time-like data of the E760 experiment 
\cite{Arm93}.
The large number of free parameters is somewhat reduced by requiring the usual 
normalization
and asymptotic conditions. Applying special nonlinear transformations 
\cite{Dub93} and incorporating the finite widths of the vector mesons, the model
describes all existing data in both the space-like and the time-like region.

In the discussion there was some criticism with regard to the large number of 
vector
mesons introduced in this model, which produces large and possibly unphysical
oscillations. Except for the possible resonance at 1.87 GeV, the higher 
resonances
are not (yet) shown by the data. However, a closer look for such resonances in 
the
time-like region by more precise experiments will be worthwhile the effort.

The contribution of strange quarks to the nucleons's structure has been
another topic of interest. Experiments to measure the strange form factors
of the nucleon by parity-violating electron scattering are being pursued
at MIT/Bates, Jefferson Lab and MAMI, and first results on the strangeness
contribution of the magnetic moment were obtained at MIT/Bates \cite{Bek97}. 
The theoretical investigations of this effect were based on dispersion 
relations, kaon-loop calculations and Skyrme models, and resulted in various
predictions differing both in sign and in magnitude. H.-W. Hammer reported on
recent studies of the
strange vector form factors of the nucleon in the framework of dispersion
relations (DR) \cite{Ham97a},
\begin{eqnarray}
\label{ImJ}
& &\bar{u}(p)\left[Im\,F^s_1(t)\gamma_{\mu}+i 
\frac{Im\,F^s_2(t)}{2m_N}
\sigma_{\mu\nu}(\bar{p}+p)^{\nu} \right]v(\bar{p})\\& &\qquad
\qquad\qquad\qquad\qquad\sim \sum_n 
\langle N(p)\bar{N}(\bar{p}) \mid n\rangle 
\langle n \mid \bar{s} \gamma_{\mu} s \mid 0 \rangle 
\,\delta^4(p+\bar{p}-p_n).\nonumber
\end{eqnarray}

The sum over the intermediate states was saturated by
$K\bar{K}$ and $3\pi$ intermediate states, which are the lightest (isoscalar) 
states containing strange and nonstrange particles, respectively. 
Due to the $1/t$ or $1/t^2$ weighting factors these states are the most important
ones in the DR. In most previous model calculations
only contributions of the kaon cloud have been included
(kaon cloud dominance), because the OZI-rule would not allow for 
matrix elements such as $\langle 3 \pi \mid \bar{s} 
\gamma_{\mu} s \mid 0 \rangle$.

For the $K\bar{K}$ continuum, 
the effects of unitarity, a realistic
kaon strangeness form factor, and the inclusion of $KN$ scattering
data were studied and compared to a
model calculation in the nonlinear $\sigma$-model \cite{Mus97,Ram97}.
The dispersion integral splits into the physical region 
($t \geq 4m^2_N$) with experimental data
for the scattering amplitude $\langle N \bar{N} \mid
K \bar{K} \rangle$ and the unphysical
region ($4m^2_K \leq t \leq 4m^2_N$) for which an analytical continuation is
needed. 
The constraints from unitarity of the $S$-matrix
for the $K \bar{K} \rightarrow N \bar{N}$ amplidudes reduce the contribution 
from the physical region by a significant
amount. As a consequence, the main contribution of the 
dispersion integral is from the unphysical region where unitarity
does not apply. As an example,the nonlinear $\sigma$-model 
violates the unitarity bounds by a factor of 3 or more.
Moreover, the impact of realistic parametrizations for
the matrix element $\langle K \bar{K} \mid \bar{s} \gamma_{\mu} s \mid 
0 \rangle$, i.e. the kaon strangeness form factor $F^s_K$, 
is significant. Comparison with the corresponding electromagnetic
quantity indicates a strong effect of the $\phi(1020)$ enhancing 
the $K\bar{K}$ threshold region in the dispersion integral.
Furthermore, the 
experimental $K^+ N$-scattering amplitudes have been analytically
continued to the $t$-channel unphysical 
region by means of backward DR \cite{Ram97}.
This technique provides the two 
$t$-channel $J=1$ partial waves that determine the spectral functions 
together with the kaon strangeness form factor $F^s_K$. 
The analytical continuation results in a resonance structure
near $K\bar{K}$ threshold, presumably due to 
the $\phi(1020)$.
In combination with $F^s_K$ this increases the 
$K\bar{K}$ contribution to the electric radius to $\rho^s_E=
-(2m^2_N/3) \left <r^2 \right >^s_E \approx 1$, which is at a level 
observable in the proposed experiments. 
Obviously the one-loop model calculations yield only a highly incomplete 
picture of the underlying physics, because they neglect the 
resonance effects in both kaon strangeness form factor and $K\bar{K}
\rightarrow N\bar{N}$ partial waves and the rescattering effects in the latter.
The predicted increase in the strangeness radius reduces the 
discrepancy between vector meson pole and $K\bar{K}$ continuum results, 
indicating a possible solution for this longstanding problem.

The $3\pi$ continuum is 
the lowest mass intermediate state in the spectral decomposition of
Eq. (10). 
Therefore its contribution is substantially enhanced in the DR in 
comparison with the kaon cloud. Since the nonresonant contribution
is very small, any sizeable effects
should be due to resonances, e.g. an $\omega$ or a $\rho\pi$ pair. 
While the treatment of the $3\pi \leftrightarrow
\omega$ resonance is similar to the pole analysis of Jaffe,
the $3\pi \leftrightarrow\rho\pi$ resonance has not been studied previously.
Because of the lack of 
experimental data for $N\bar{N}\to \rho\pi$,
the corresponding amplitude was calculated in the Born approximation
of the linear $\sigma$-model. The transition form factor
$\langle \rho \pi \mid \bar{s} \gamma_{\mu} s \mid 0 \rangle$ was taken from
$\phi$-dominance and data for the $\phi\to\rho\pi$ decay which also violate
the OZI-rule.
The results indicate that, against naive expectations, resonance
effects can enlarge the $3\pi$ contribution to a level comparable
with typical kaon cloud contributions \cite{Ham97b}. Therefore, the assumption
of kaon cloud dominance, which is based on the OZI-rule, may be questionable.

Following earlier investigations of the pion form factors \cite{Gas91},
T. Hannah calculated these observables
by means of a dispersion relation for the inverse of the form factor
using exact unitarity and the chiral expansion. This inverse amplitude method
(IAM) has been
applied to both one \cite{Tru88} and two loops \cite{Han97} in
the chiral expansion with results that are formally equivalent to the
[0,1] and [0,2] Pad\'{e} approximants, respectively. Since the expansion
of the inverse of the form factor to one loop in ChPT is equivalent to
the standard vector meson dominance models \cite{Dun96}, the IAM to
two loops contains the additional two-loop chiral corrections to these
models.

The vector form factor $F_V$ is experimentally well known both in the
time-like region from electron-positron annihilation and in the space-like
region from $e\pi$ scattering and electroproduction of pions. In
addition, the isovector part of the form factor has recently been
determined very precisely from the decay
$\tau\rightarrow\pi\pi\nu_{\tau}$ using conservation of the vector current.
From these experimental data some of the low-energy constants in the IAM
to two loops have been determined and the rest has been fixed from
$\pi\pi$ scattering. As for the IAM to one loop, the low-energy constant
$\bar{l}_6$ has been determined from the experimental value of the
electromagnetic radius of the pion, and finally for ChPT the low-energy
constants have been determined in the same way as in Ref.
\cite{Gas91}.

The results of the IAM to two loops
agree nicely with the experimental data over a large
energy region, and even the IAM to one loop describes the data
qualitatively in the same energy region (see Fig. 6). In addition, 
the IAM has been
analyzed in the complex energy plane and one finds the pole on the second
Riemann sheet corresponding to the $\rho$(770) resonance. Thus, in the case of
the vector form factor, the IAM is a very successful way of extending
the range of applicability of ChPT order by order in the chiral
expansion.

\begin{figure}[h,t,b]
\begin{center}
\epsfig{file=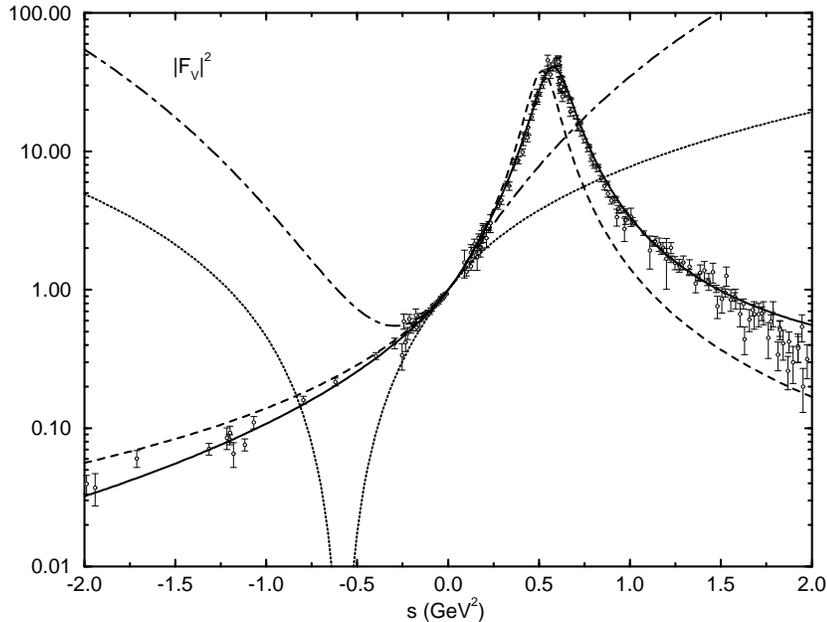, width=8cm, angle=270}
\caption{\small The pion vector form factor $\mid F \mid^ 2$ as function
of $s = q^2$ in
the space- and time-like regions. The solid line is the IAM to two loops, the 
dashed line the IAM to one loop, the dash-dotted line two-loop ChPT, and the
dotted line one-loop ChPT.}
\end{center}
\end{figure}

\section{Hadron Polarizabilities}

The field of hadron polarizabilities was reviewed in three plenary
talks at this Workshop, B. Holstein \cite{Hol97}
reported on ChPT calculations, N.d'Hose \cite{dHo97} on
experiments to measure the polarizabilities of the nucleon, and 
M. Moinester \cite{Moi97} on 
hadron polarizability experiments at Fermilab and CERN. In the Working
Group, V. Olmos reported on the status of the proton polarizability
and an ongoing experiment at Mainz. According to the low energy 
theorem \cite{Gel54, Pet81} the
differential cross section for Compton scattering can be cast into the form

\begin{equation}
 \frac{d\sigma}{d\Omega} = \left (\frac{d\sigma}{d\Omega} \right )_P
  - r_0(\frac{\omega'}{\omega})^2\omega \omega' 
  \left [\frac{\alpha+\beta}{2}(1+\cos \theta)^2
  +\frac{\alpha-\beta}{2}(1-\cos \theta)^2 \right ] + ... \, , 
\end{equation}

with $r_0 = e^2/(4 \pi m_N)$ the classical proton radius. The first term on the $rhs$
of eq. (11) is the Powell cross section describing the scattering off a 
point particle with an anomalous magnetic moment. It can be expressed by a
somewhat lengthy formula in terms of the global properties of the nucleon,
its charge $Q_N$, mass $m_N$ and anomalous magnetic moment $\kappa_N$.
The second term is of quadratic order in the photon energies $\omega$ and
$\omega'$. It contains
the structure effects in the form of the electic $(\alpha)$ and magnetic
$(\beta)$ polarizabilities.
Fig. 7 shows the relative importance of the different contributions to the
cross section. The nonrelativistic (``Thomson'') and relativistic (``Klein-
Nishina'') results for scattering on an ideal point particle are quite similar
but differ dramatically from the Powell cross section \cite{Pow49} 
precisely because of the
anomalous magnetic moment. However, this increase of the cross section due to
the anomalous magnetic moment is largely compensated by the effect of the
polarizabilities and the $\pi°$ pole term (Wess-Zumino-Witten term, anomaly).
Of course, the anomalous magnetic moment, the polarizabilities due to excited
intermediate states and the anomaly are different aspects of the same fact,
namely the compositeness of the nucleon. Somehow these 3 effects conspire
to give away only little information about the internal structure as 
contained in the terms $O(\omega^2)$, as may be seen from the curve labeled
"LET". 
The steep rise in the curve labeled "L'vov"   \cite{Lvo97a}
is due to higher order terms in $\omega$, calculated by a dispersion calculation
treating the $\Delta$ resonance and other high energetic effects.

The optical theorem and dispersion relations lead to Baldin's sum rule 
\cite{Bal60},

\begin{equation}
 \alpha_p+\beta_p = \frac{1}{2\pi^2} \int^{\infty}_{thr} \frac
 {\sigma_{tot}(\omega)}{\omega^2} d \omega = (14.2 \pm 0.5) 10^{-4}fm^3,
\end{equation}

i.e. the sum of the two polarizabilities is well determined by the total
photoabsorption cross section $\sigma_{tot}$. As may be seen from eq. (11),
the sum of the polarizabilities enters for forward scattering $(\theta = 0)$
, the difference for backward scattering $(\theta = \pi)$,
and for $\theta = \pi/2$ only $\alpha$ contributes. In spite of considerable
experimental efforts, $\alpha$ and $\beta$ have not yet been determined
independently by Compton scattering to a good accuracy. The present result
(in units of $10^{-4}fm^3)$ is mostly due to the work of the Illinois/SAL
collaboration \cite{Hal93,Mac95}, 
$\alpha_p = 12.1 \pm 0.8 \pm 0.5$, $\beta_p = 2.1 \mp 0.8 \mp 0.5$,
the change of signs indicating that Baldin's sum rule has been used as a
constraint.

\begin{figure}[ht]
\begin{minipage}[ht]{160mm}
\begin{center}
\epsfig{file=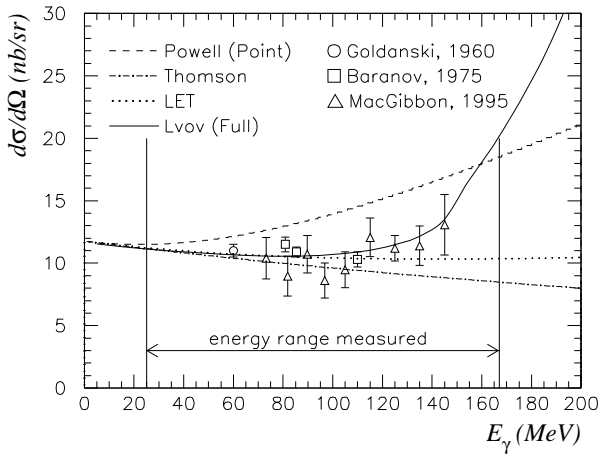, height = 6cm}
\caption{Differential cross section for Compton scattering
         off the proton at $\theta = 90^{\circ}$. See text}
\end{center}
\end{minipage}
\end{figure}

The  experiment reported by V. Olmos was performed with 
tagged photons from the Glasgow Tagger \cite{Ant91} at the minimum 
available electron energy from MAMI, i.e. 180 MeV.
With this setup photon energies between 25 and 167 MeV could be tagged
with an energy resolution of 0.5 MeV and a tagging efficiency of about
20\%\  .
The scattered photons were detected with the TAPS photon spectrometer 
\cite{Nov91}
consisting of 384 $BaF_{2}$ crystals plus plastic anti-charged particle
detectors. 
The TAPS-tagger time resolution is 1.4 ns FWHM.
The energy resolution of TAPS is 14.5 MeV.
TAPS allowed to  measure at polar angles between $25^{\circ}$ and 
$165^{\circ}$.
Due to the strong collimation and the smallness of the cross section to be
measured, there are backgrounds from
electromagnetic processes at the collimators. 
These give an important
contribution at low energies (25--50 MeV) and small angles 
($25^{\circ}-60^{\circ}$).
Other sources of background are cosmic rays that lead to random 
coincidences with the tagger. 
In this case a suppression  is possible with
a time cut. The recently taken Mainz data are now being analyzed with the promise to 
determine $\alpha_p$ and $\beta_p$ independently with an error bar of 
$10^{-4} fm^{3}$. 

Information on the neutron is much more
 difficult to obtain. While a value $\alpha_n = (12.0 \pm 1.5 \pm 2.0)$
was reported from neutron scattering off heavy nuclei \cite{Sch91}, a
more recent investigation comes to a completely different result with 
large error bars, 
$ \alpha_{n} = (0\pm5)$ \cite{Koe95}.
The problem with Compton scattering is that the Thomson term, and hence 
its interference with the
polarizability term, vanishes for the neutron. Therefore the leading neutron
contribution at low energies, 
$\sigma_n \sim (\alpha_n \omega^2)^2$, is difficult to 
disentangle
from the much larger proton contributions in nuclei. Though there have been
efforts to determine $\alpha_n$ in quasifree kinematics \cite{Ros90}, 
the effects of final state interactions and meson exchange currents have to be
carefully analyzed.

As an alternative D. Hornidge presented data on a recent Saskatoon experiment
on coherent photon scattering off the deuteron. 
The experimental difficulty arises from the necessity of resolving the elastic
peak from the inelastic contribution. Up to now, there was only one measurement
on this reaction (at Illinois \cite{Luc94} below 70 MeV). Using the photon
tagging spectrometer and the high resolution Boston University NaI detector
at the Saskatchewan Accelerator Laboratory (SAL), a measurement of elastic 
photon scattering from deuterium was accomplished. Tagged photons in the energy
range $E_{\gamma} = $ 84.2 - 104.5 MeV were scattered from a liquid deuterium
target and preliminary differential cross sections were measured at five angles
covering the range $\theta_{\gamma'} = 35^{\circ} - 150^{\circ}$.
In a somewhat symbolical way,
the cross section for this reaction is $\sigma_d \sim (f_{p,Th} - \alpha_p \omega^2
- \alpha_n\omega^2)^2$, i.e. the isoscalar sum $\alpha_p + \alpha_n$ interferes
with the Thomson term of the proton, leading to a considerable enhancement of
the neutron effect. However, the analysis would still require a more rigorous
calculation of meson exchange and isobar currents in the deuteron, in order to
obtain precise values for the neutron's polarizabilities. 
Complex calculations by Levchuk \cite{Lev95} and Wilbois \cite{Wil95} attempt
to explain these effects. The statistics of this SAL measurement are such
that the errors in the differential cross sections are small enough
(5 \% - 10 \%) to show that neither the Levchuk calculation (with a wide
range of polarizabilities) nor the Wilbois calculation agree with the
preliminary data.

By means of polarization experiments it will be possible to also 
measure the 4 spin
(or vector) polarizabilities of the proton. As shown by T. Hemmert, the
amplitude for Compton scattering off a proton $(\;\gamma
\;p\rightarrow \gamma^{\prime} \; p^\prime \;)$ can be written in terms 
of 6 structure functions $A_i(\omega , \theta )$, 

\begin{eqnarray}
T &=& A_1(\omega,\theta)\;\vec{\epsilon}^{* \prime}\cdot\vec{\epsilon}  
   +A_2(\omega,\theta)\;\vec{\epsilon}^{* \prime}\cdot\hat{k} \; \vec{\epsilon}
     \cdot\hat{k}^\prime 
  +A_3(\omega,\theta)\;i\vec{\sigma}\cdot(\vec{\epsilon}
^{* \prime}\times
     \vec{\epsilon})\\ \nonumber  
&& +A_4(\omega,\theta)\;i\vec{\sigma}\cdot(\hat{k}^\prime \times\hat{k})
     \vec{\epsilon}^{* \prime} \cdot\vec{\epsilon} 
  +A_5(\omega,\theta)\;i\vec{\sigma}\cdot[(\vec{\epsilon}^{* \prime} \times
     \hat{k}) \vec{\epsilon}\cdot\hat{k}^\prime -(\vec{\epsilon}\times
     \hat{k}^\prime ) \vec{\epsilon}^{* \prime} \cdot\hat{k}] \\ \nonumber
&&
   +A_6(\omega,\theta)\;i\vec{\sigma}\cdot[(\vec{\epsilon}^{* \prime}\times
     \hat{k}^\prime ) \hat{\epsilon}\cdot\hat{k}^\prime -(\vec{\epsilon}\times
     \hat{k})\vec{\epsilon}^{* \prime} \cdot\hat{k}].
\end{eqnarray}

Here $\vec{\epsilon},\hat{k}\; (\vec{\epsilon}^{\prime} ,\hat{k}^{\prime} )$ 
are 
the polarization vector and the direction  of the incident (final) photon, while 
$\vec{\sigma}$ denotes the spin vector of the nucleon. The structure-dependent
polarizabilities are of relative order $\omega$ with regard to the leading
terms of a Taylor series in $\omega$,  
  
\begin{eqnarray}
A_1(\omega,\theta) &=& -\frac{e^2}{m_p}+4\pi \left( \alpha+
  \cos \theta \right) \omega^2 
  -\frac{e^2}{4m_p^3} \left(1-\cos \theta\right) \omega^2 + \ldots \nonumber \\
A_2(\omega,\theta) &=& \frac{e^2}{m_p^2} \omega-4\pi\beta
  \omega^2 +\ldots\nonumber\\
A_3(\omega,\theta) &=& \left[1+2 \kappa_p - (1+\kappa_p)^2 \cos \theta\right]
  \frac{e^2}{2 m_p^2} \omega
 -\frac{(2\kappa_{p}+1) e^2}{8m_p^4} \cos \theta \; \omega^3 \nonumber \\
 & &+ 4\pi \left[ \gamma_1 -(\gamma_{2} + 2\gamma_{4}) \cos \theta \right]
   \omega^3 +\ldots \nonumber\\
A_4(\omega,\theta) &=& -\frac{(1+\kappa_p)^2 e^2}{m_p^2}\omega + 4\pi
 \gamma_{2} \omega^3+\ldots\nonumber\\
A_5(\omega,\theta) &=& \frac{(1+\kappa_p)^2e^2}{2m_p^2}\omega+4\pi
 \gamma_{4} \omega^3 +\ldots\nonumber\\
A_6(\omega,\theta) &=& -\frac{(1+\kappa_p)e^2}{2 m_p^2}\omega+4\pi
 \gamma_{3} \omega^3 +\ldots, 
\end{eqnarray}

\noindent
which defines the scalar polarizabilities $\alpha$ and $\beta$, and the vector
polarizabilities $\gamma_1$, $\gamma_2$, $\gamma_3$ and $\gamma_4$ of the 
proton.
The latter were calculated by Bernard et al. \cite{Ber95} in HBChPT at
$O(p^3)$. Hemmert et al. \cite{Hem97} have now explicitly included $\Delta$ 
degrees
of freedom in an expansion to $O (\epsilon^3)$, where $\epsilon$ denotes an
external momentum, a quark mass, or the N-$\Delta$ mass splitting. The calculation
includes the anomaly, $\Delta$ pole terms and both N-$\pi$ and $\Delta-\pi$ loops, whose contributions are shown in Tab. 1. 

\bigskip
\begin{center}
\begin{tabular}{c|rrrr|r|r|l}
             & $N\pi$  & WZW & $\Delta$  & $\Delta\pi$ & $O(\epsilon^3)$ & $O(p^3)$ 
             & ``experiment'' \\
\hline
$\gamma_{1}$ & $+ 4.6$ & $- 22.0$ & $0$  & $- 0.21$    & $- 17.6$       & $-17.4$
             & \\
$\gamma_{2}$ & $+ 2.3$ & $0$   & $- 2.4$ & $- 0.23$ &    $- 0.3$        & $+2.3$
             &\\
$\gamma_{3}$ & $+ 1.2$ & $+ 11.0$ & $0$ & $- 0.12$ &     $+ 12.1$       & $+12.2$
             & \\
$\gamma_{4}$ & $- 1.2$ & $- 11.0$ & $+ 2.4$ & $+ 0.12$ & $- 9.7$        & $-12.2$
             &\\
\hline
$\gamma_{0}$ & $+ 4.7$ & $0$ & $-2.4$ & $- 0.2$          & $+ 2.0$      & $4.7$
             & $-1.34$ \\
\hline
$\gamma_{\pi}$ & $+ 4.5$ & $-44.0$ & $+ 2.4$ & $- 0.2$   & $-37.2$      & $-39.5$
             & $-(27.6\pm1.9\pm3.2)$ \\

\end{tabular}
\vspace{0.5cm}
\end{center}
Table 1: The vector polarizabilities of the proton as calculated in 
HBChPT to $O(\epsilon^3)$ by Hemmert et al. \cite{Hem97} and to $O(p^3)$ 
by Bernard et al.
\cite{Ber95}. For the individual contributions see the text.
The results are given in units of $10^{-4}\;fm^4$.

\vspace{0.5cm}
Is is seen from the table that the anomaly (Wess-Zumino-Witten term, WZW)
dominates 3 of the vector polarizabilities. Only the numerically small
$\gamma_2$ is free of a WZW contribution. In this case the N$\pi$ loops are
predicted to be canceled by $\Delta$ contributions, leading to a change of
sign in the net result. Also the value of $\gamma_4$ is very sensitive to
$\Delta$ degrees of freedom. The forward scattering amplitude 
is determined by the structure functions $A_1$ and $A_3$ (see eq. (13)). 
According to eq. (14) the spin-independent cross section is then a function
of $\alpha+\beta$, while the spin-dependent one measures the combination
$\gamma_1-\gamma_2-2\gamma_4 = \gamma_0$. The former combination is
known from Baldin's sum rule, the latter one is related to an integral over
the difference of the absorption cross sections for the helicity states 3/2
and 1/2, similar to the Gerasimov-Drell-Hearn sum rule (GDH), but weighted with
an additional
$\omega^{-2}$. As may be seen in Tab. 1, the experimental value \cite{San94}
differs from
the predictions even in sign. While the forward spin polarizability $\gamma_0$
is small and independent of the anomaly, the backward spin polarizability
$\gamma_{\pi} = \gamma_1+\gamma_2+2\gamma_4$ (see eq. (14)!) is very large,
essentially because of the anomaly contribution. The corresponding experimental
value as estimated by a multipole analysis \cite{LEG97} is shown in Tab. 1 and compared
to the predictions. When comparing theory and experiment, it has to be kept
in mind that the data were determined indirectly from a multipole analysis
and/or extrapolation of unpolarized photon scattering. The full set of spin
polarizabilities could be extracted by scattering circularly polarized photons
off polarized nucleons, which will be another important test of low energy
QCD and ChPT. On the theoretical side it will be necessary to study
$O(\epsilon^4)$ corrections to the results given in Tab. 1 in order to
check the convergence of the perturbation series.

\def\Re{\mbox{\,Re\,}}
\def\Im{\mbox{\,Im\,}}

Concerning a somewhat related object, A. L'vov reported on a test of the GDH 
\cite{GDH65},
which relates a weighted integral of the spin-dependent photoabsorption
cross section $\Delta\sigma=\sigma_{3/2}-\sigma_{1/2}$ to the
anomalous magnetic moment of the nucleon.
A failure to saturate the GDH for the isospin-odd channel
may imply \cite{San94,Lvov97b} either a large cross
section $\Delta\sigma(\gamma n) \simeq -\Delta\sigma(\gamma p) \sim
50\,\mu$b in the few-GeV energy region or, alternatively, a violation of
the Vector Meson Dominance for the spin-dependent transitions at high
energies.  Presently, an experiment at Mainz is scheduled to measure
$\Delta\sigma(\gamma p)$ up to 800 MeV using a large
$4\pi$ detector. Further measurements up to a few GeV will be
necessary to establish whether the GDH integral converges
to its canonical value.

It was proposed \cite{Lvov97b} to determine the forward Compton scattering
amplitude,
\begin{equation}
   f(\omega) = \vec\epsilon^{\;\prime}\!\cdot\vec\epsilon\, f_1(\omega)
 + i\omega\vec\sigma\cdot\vec\epsilon^{\;\prime}\!\times\vec\epsilon\,
     f_2(\omega),
\end{equation}
from the reaction $\gamma p \to e^+e^- p$ by studying
the interference between the Bethe-Heitler (BH) and the virtual Compton
Scattering (VCS) amplitudes of the reaction. Since the BH mechanism
produces $e^+e^-$ pairs in states of positive C-parity and the VCS
gives C-odd pairs, such an interference results in an asymmetry between
$e^+$- and $e^-$-yields in the same kinematics.  In the regime where the
transverse momentum $Q$ of the pair and its invariant mass $M_{e^+e^-}$
are small, the VCS amplitude is determined by the same functions $f_1$
and $f_2$ which describe the forward real Compton scattering.  The
corresponding asymmetry \cite{Lvov97b}, 
\begin{eqnarray}
   \Sigma(e^+/e^-) & \sim &
  - \frac{137\, Q^3}{\omega M_{e^+e^-}} \Big\{
 (\Re f_1(\omega)-h_\gamma h_N \omega \Re f_2(\omega)) \cos\phi
  \nonumber\\ && \hspace{4em} {} +
 (-h_\gamma \Im f_1(\omega)+ h_N \omega \Im f_2(\omega)) \sin\phi \Big\},
\end{eqnarray}
depends on the photon helicities $h_\gamma=\pm 1$ and the proton spin
projections $\frac12 h_N = \pm\frac12$ upon the beam direction and
determines a preferred azimuthal orientation $\phi$ of the pair (i.e.
the $e^+{-}e^-$ plane) with respect to its transverse momentum
$\vec Q$.  Measuring the asymmetry $\Sigma(e^+/e^-)$ on a polarized
target, one can find the imaginary part of $f_2$ and hence
$\Delta\sigma=-8\pi\Im f_2$, according to unitarity.  Moreover,
with circularly polarized photons one can also find $\Re f_2$, which is
related to $\Im f_2$ by a dispersion relation and gives further
constraints for $\Im f_2$.
Since the GDH sum rule is valid if and only if the function $f_2$
vanishes at high energies, direct measurements of $f_2$ in the 
reaction $\gamma p \to e^+e^- p$ at a few GeV 
could serve as a sensitive test of the GDH sum rule.

Our present knowledge about the pion polarizability is even more 
unsatisfactory than in the case of the nucleon.
In the past, information has been obtained from radiative pion scattering off
heavy nuclei at Serpukhov $(\alpha _{\pi\pm} = 6.8 \pm 1.4 \pm 1.2)$, from
radiative pion photoproduction $\gamma p \rightarrow \gamma\pi^+n$ at the
Lebedev Institute with the result $20 \pm 12$, and from various evaluations
of the reaction $\gamma\gamma \rightarrow \pi^+\pi^-$. The latter results are
quite sensitive to badly known off-shell effects and lead to a wide range of
predictions.
The value of the polarizability derived from radiative pion photoproduction 
\cite{Aib86} has
very large error bars and is at variance with ChPT.
It has therefore been proposed to repeat this experiment at MAMI.
 
The differential cross section of elastic $\gamma \pi^+$ scattering
and the difference of the polarizabilities
$\alpha_{\pi^+} -\beta_{\pi^+}$ can be found by extrapolating the
experimental data on radiative pion photoproduction 
to the pion pole \cite{Aib86, Dre94}. An appropriate kinematics for 
the experiment was suggested \cite{Ahr95} in order to increase 
the contribution of the polarizability  in the physical region, 
to avoid additional singularities close to the region of extrapolation 
and to separate the signal from the 
background processes. The calculation takes 
account of the $P_{33}(1232)$ and $P_{11}(1440)$ resonances and shows that in 
the suggested kinematical region the contribution of resonances should be
under control. As a result of extrapolation at fixed $s_1$ (energy
of the pion-photon final state),  should not depend on 
the incident photon energy $\nu_1$, which can be used to
check the correctness of the extrapolation 
procedure. The simultaneous analysis for all values of $s_1$ and $\nu_1$ collects
information  from the full physical region and decreases 
the error obtained by the extrapolation procedure. For the triple coincidences 
and an exposition time of 60 days, the precision is expected to be
$\Delta \alpha_{\pi_+}=1.3$.

\section{Virtual Compton Scattering}

The process of virtual Compton scattering (VCS) off the proton can be 
observed by the reactions $e^- p \rightarrow e^- p \gamma$ and $\gamma p 
\rightarrow p e^+ e^-$. In both cases, there appears a strong background
of BH scattering together with the VCS signal. In the case of pair
production the intermediate photon is time-like $(q^2 > 0)$, at low 
energies this process determines the polarizability of the nucleon 
in the time-like region. A possible application of this reaction in the
limit of $q^2 \rightarrow 0$ to test the GDH sum rule, was discussed
above. In the following we will only address radiative electron 
scattering, which tests the polarizability in the space-like region
$(q^2 < 0)$.

At low energies the amplitude can be expanded in a power series in $\omega'$,
the energy of the emitted photon. The terms in $(\omega')^{-1}$ and
$(\omega')^{0}$ may be expressed by the global properties of the nucleon,
i.e. charge, mass, anomalous magnetic moment and the elastic form factors
$G_E$ and $G_M$ \cite{Sch96}. The higher terms depend
on the internal structure of the nucleon. To leading order in $\omega'$,
there appear 10 generalized polarizabilities (GPs), as was shown by
Guichon \cite{Gui95}. The restriction to terms linear in $\omega'$ corresponds
to the dipole approximation, i.e. the outgoing real photons have
$E1$ or $M1$ radiation. The selection rules of parity
and angular momentum then result in 10 GPs, 3 scalar and 7 vector (or spin)
polarizabilities. However, the additional constraint of $C$ parity  and
nucleon crossing symmetry reduces this number to 6 independent GPs to
that order, 2 scalar and 4 vector ones \cite{Met97, Dre97a, Dre97b}.

S.~Scherer reported on  a calculation of the GPs
within the heavy-baryon formulation of chiral perturbation theory (HBChPT) 
to third order in the external momenta.
   At $O(p^3)$, contributions to the GPs are generated by nine one-loop 
diagrams and the $\pi^0$-exchange $t$-channel pole graph \cite{Hem97a}.   
   For the loop diagrams only the leading-order Lagrangians,
$\widehat{\cal L}_{\pi N}^{(1)}$ and ${\cal{L}}_{\pi \pi}^{(2)}$
are needed \cite{Eck96}.
   The $\pi^0$-exchange diagram involves  the $\pi^0\gamma\gamma^\ast$ vertex 
provided by the Wess-Zumino-Witten Lagrangian.
   Some numerical results for the GPs are shown in Fig. 8.
   At $O(p^3)$, the results only depend on the pion mass $m_\pi$, the axial 
coupling constant $g_A$, and the pion decay constant $F_\pi$.

\begin{figure}[h,t,b]
\begin{center}
\epsfig{file=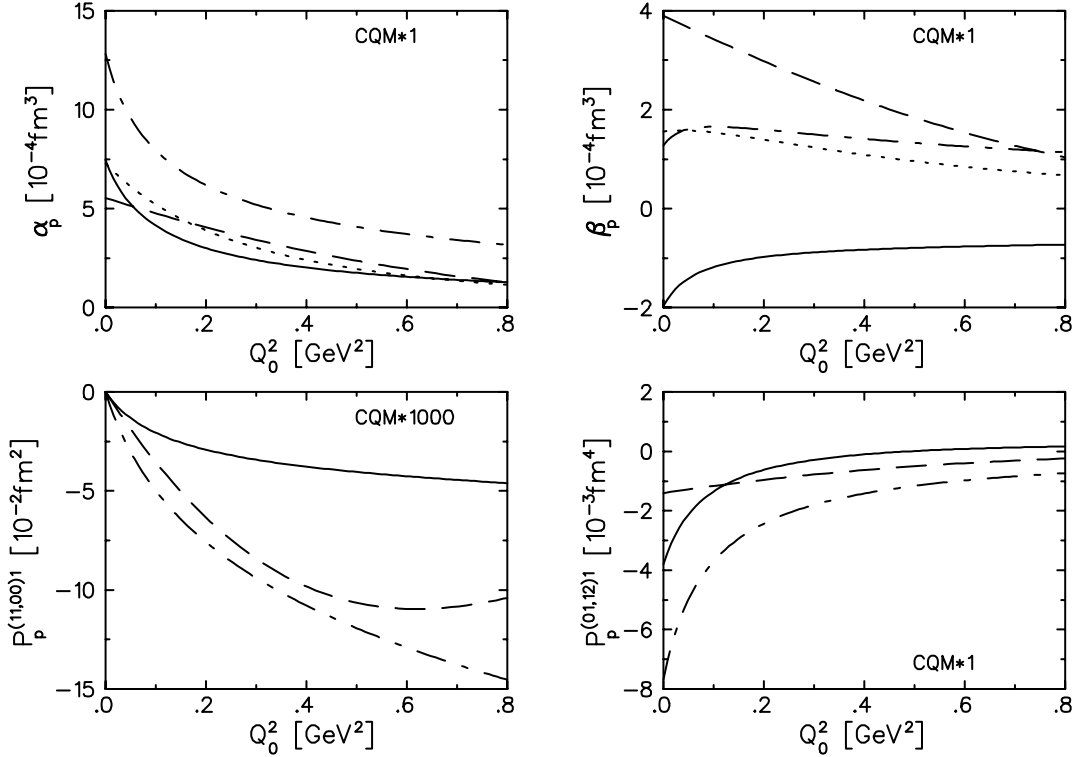, height=10cm}
\caption{\small 
Generalized polarizabilities of the proton. Solid line:
calculation with the linear sigma model \cite{Met97}, dashed line:
constituent quark model \cite{Gui95}, dash-dotted line: HBChPT \cite{Hem97a},
dotted line: effective lagrangian model \cite{Van96}. The results of the 
quark model have been scaled.}
\end{center}
\end{figure}

For example, the prediction for the generalized electric polarizability 
 of the proton, 
$$
\frac{\alpha_p({q}^2)}{\alpha_p(0)}=
1-\frac{7}{50}\frac{q^2}{m^2_\pi}
+\frac{81}{2800}\frac{q^4}{m^4_\pi}
+O\left(\frac{q^6}{m^6_\pi}\right),\,\,
\alpha_p=\frac{5 e^2 g_A^2}{384\pi^2m_\pi F_\pi^2}
=12.8\times 10^{-4}\,\mbox{fm}^3,
$$
decreases considerably faster with $q^2$ than in the constituent quark 
model \cite{Gui95}.
    The predictions for the spin-dependent GPs originate from two rather 
distinct sources---an isoscalar piece from pionic loop contributions,
and an isovector piece from the $\pi^0$ exchange diagram
contributing to the spin-dependent GPs only.
   The loop contribution to the spin-dependent GPs is generally
much smaller than the the $\pi^0$-exchange contribution. 
   For example, the spin-dependent GP $P^{(11,11)1}$ is given by
$$
P^{(11,11)1}(q^2)=-\frac{1}{288}\frac{g_A^2}{F_\pi^2}
\frac{1}{\pi^2 M}\left[\frac{q^2}{m^2_\pi}-\frac{1}{10}
\frac{q^4}{m^4_\pi}\right]
+\frac{1}{3M}\frac{g_A}{8\pi^2F^2_\pi}\tau_3
\left[\frac{q^2}{m^2_\pi}-\frac{q^4}{m^4_\pi}\right]
 +O\left(\frac{q^6}{m^6_\pi}\right),
$$
   and vanishes at the origin as a consequence of $C$ invariance \cite{Dre97b}.
   This is not true for the prediction in the constituent quark model 
   \cite{Gui95}.
It is also worthwhile pointing out that for 5 of the 7 vector GPs,
HBChPT results in much larger values than in the case of the quark model.

According to Guichon \cite{Gui95}, the GPs are denoted by $P^{(\rho' L',
\rho L) S}$, with $\rho = $0,1 and 2 for Coulomb, magnetic and electric
multipoles, and $L$ standing for the multipolarities. The superscript
$S$ can take the values 0 (scalar GPs) and 1 (vector or spin GPs).
Since the electric $(\rho = 2)$ and Coulomb $(\rho = 0)$ multipoles
are related in the low energy limit by gauge invariance (Siegert's theorem),
Guichon expressed the electric transitions $(\rho = 2)$ by the Coulomb
$(\rho = 0)$ ones plus a remainder, which leads to so-called mixed
multipoles $\hat{P}^{(\rho' L', L) S}$. A. Metz explained in detail
how the 10 GPs defined in this way are related among themselves and with the
polarizabilities for real photons obtained at $q = 0$
\cite{Dre97a, Dre97b}. In particular, the 
3 scalar GPs $P^{(01,01)0}, P^{(11,11)0}$ and $\hat{P}^{(01,0)0}$
are related such that $\hat{P}^{(01,0)0}$ can be eliminated.
In the real photon limit, $q \rightarrow 0, P^{(01,01)0} \sim \alpha$
and $P^{(11,11)0} \sim \beta$. At finite $q$, these GPs probe
$\alpha (q^{2})$ and $\beta (q^2)$, which are related to
the spatial distributions of the electric and magnetic dipole moments in the
nucleon.

The 7 vector GPs $P^{(01,01)1}, P^{(11,11)1}, P^{(11,00)1},
P^{(01, 12)1}, P^{(11, 02)1},
\hat{P}^{(11,2)1}$ and $\hat{P}^{(01,1)1}$
are related by 3 equations such that the 2 "mixed" GPs and one further GP can 
be eliminated. In the limit of $q \rightarrow 0,\; P^{(01,12)1} \sim \gamma_3$
and $P^{(11,02)1} \sim (\gamma_2 + \gamma_4)$, describing the Coulomb 
dipole/magnetic quadrupole and Coulomb quadrupole/magnetic dipole 
interferences, respectively. The other 3 GPs, corresponding to 
$C1/C1, M1/M1$ and
$M1/C0$ radiation in order, vanish in the limit $q \rightarrow 0$.

A.~Metz further reported on calculations of the GPs in the linear sigma
model (LSM) \cite{Met97}. Since this model fulfills all the necessary symmetry relations
and because of its simplicity, the relations between the GPs were first found
in the framework of that model. 
In fact, the leading terms of the polarizabilities
in a power series expansion in $\mu = m_{\pi}/m_N$ agree with the 
heavy baryon ChPT \cite{Hem97a}.

In the differential cross section for unpolarized particles, the interference
term between BH and VCS contains 4 structure functions $P_{LL}, P_{TT},
P_{LT}$ and $P'_{LT}$, which can be separated by a super-Rosenbluth plot.
Due to the relations between the GPs, however, the latter two functions
are equivalent up to a kinematical factor. Therefore, such an experiment
will provide information on 3 (combinations) of GPs only. As was recently
shown by M.~Vanderhaeghen \cite{Van97}, the additional information may be
obtained in the reaction $p (\vec{e}, e' \vec{p}) \gamma$, by measuring
the asymmetries for different directions of the recoil nucleon polarization.
This would lead to 3 further structure functions $P^z_{LT}, P^{'z}_{LT},
P^{'x}_{LT}$, with $x$ and $z$ corresponding to nucleon polarization in
the hadronic plane, perpendicular and parallel to the virtual
photon, respectively. In the calculation of Metz, the GPs lead to a decrease
of the cross section for a typical MAMI kinematics by about $2-10 \%$,
increasing with $\omega'$. In the case of the asymmetries, the effect of
the GPs can have both signs, with a somewhat smaller absolute value
$(2-6 \%)$. In view of the unexpectedly large radiative corrections
and the smallness of the effects, a full Rosenbluth separation will
be quite a challenge.

VCS is presently being investigated at MIT-Bates (proposal 97-03), MAMI
(collaboration A1) and Jefferson Lab \cite{dHo97}. It is the aim of the former two
experiments to measure the GPs of the proton, while there are also proposals 
to study the resonance region and the quark distribution of the nucleon at
Jefferson Lab.

R. Miskimen reported on the Bates proposal which has been optimized 
for measurements of $\alpha (q)$ and $\beta (q)$ at low momentum transfer.
The scattered protons will be detected in a cluster of three OOPS modules 
(Out-Of-Plane-Spectrometers), two of which will be placed out of the 
scattering plane. Out-of-plane detection is an important capability for 
VCS experiments.  By going above or below the scattering plane, the BH
process can be suppressed relative to VCS
because one can move away from the direction of the incident 
and scattered electrons. By going out of plane it is also possible to 
reach low values of momentum transfer, where the kinematical focussing 
of the proton weakens. Clearly data at low momentum transfer are needed 
in order to test ChPT. 

Data will be simultaneously taken using three OOPS modules at 3 combinations
of scattering angles, $(\theta, \phi) = (90^{\circ}, 90^{\circ}), (90^{\circ},
270^{\circ})$, and $(90^{\circ}, 180^{\circ})$. For the first two kinematical
conditions the experiment is sensitive to $(\epsilon P_{LL} - P_{TT})$, in the
last case also $P_{LT}$ is being probed. Data will be taken with $\epsilon =
0.9$ to maximize the sensitivity 
to $P_{LL}$, which is proportional to $\alpha (q)$.  The initial photon 
momentum 
will be 240 MeV, or $Q^2 \approx 0.05 GeV^2$, which is in the 
anticipated range of validity of chiral perturbation theory.  
Because the electron scattering virtual photon flux factor is 
approximately 100 times larger than typical real photon tagging rates, 
the extracted polarizabilities should have greatly reduced statistical 
uncertainty compared to the real Compton scattering case. Taking VCS data 
simultaneously will also reduce the  systematic errors of the experiment. 

J.~Roche presented some preliminary results of the VCS collaboration at MAMI.
Data have been taken in the scattering plane at $Q^2 = 0.33 GeV^2$ and 
virtual photon polarization $\epsilon = 0.62$. As is shown in Fig. 9,
forward scattering is completely dominated by the BH process such that VCS can only be 
studied under backward angles. The prediction of the low energy expansion 
(full curve in Fig. 9) is verified within a precision of $10 \%$. It includes
the BH background, the Born terms and radiative corrections to BH scattering.
The size of the radiative corrections turned out to be quite large, it is of
the order of the structure effects to be measured. At the low final photon
energy $(\omega' = q' = 45 MeV)$, such effects of the GPs will be
very small, while $10-15 \%$ effects are expected at larger energies, e.g.
$\omega' = 90 MeV$. The goal of the experiments is to reach statistical and
systematical errors of less than $3 \%$. In order to achieve this
accuracy further improvements are necessary with regard to the stability 
between the runs and the global normalization of the data. However, this
pilot experiment has clearly demonstrated that measurements of the GPs are
within reach. Concerning the necessary accuracy of the data, it should be
kept in mind that the dependence of the GPs on the momentum transfer $Q^2$
is predicted quite differently in various models. In particular pion cloud
contributions lead to large transition radii and GPs decreasing much more
rapidly with $Q^2$ than in the case of quark and phenomenological
models. Similarly the low energy constants introduced in ChPT correspond
to point interactions, i.e. form factors are only build up by higher loop
corrections. This will also be the case for the triangle anomaly (WZW term, 
$\pi^{\circ}$ exchange) which is responsible for the large values of some of 
the vector polarizibilities. The internal (spatial) structure of this
term could be resolved by VCS with polarization degrees of freedom.

\begin{figure}[h]
\begin{center}
\epsfig{file=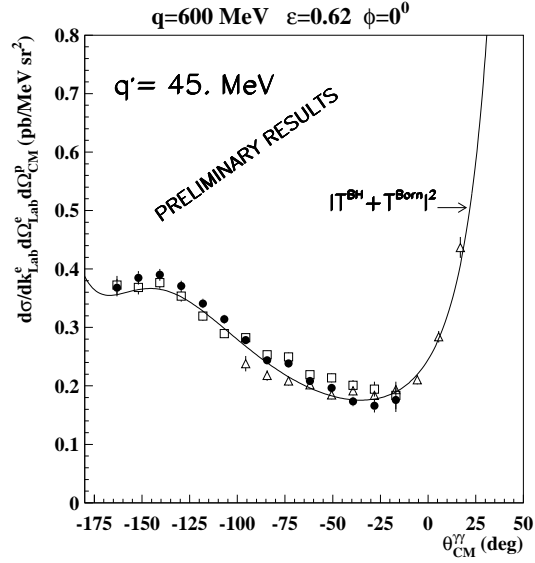, width=7cm}
\caption{\small The differential cross section for virtual Compton
scattering at $\omega' = q' = 45 MeV$. Preliminary results of the VCS
collaboration at MAMI. See text and Ref. \cite{dHo97}.}
\end{center}
\end{figure}

\section{Acknowledgement}

It is a pleasure to thank all the participants of the working group for their
contributions and the lively discussions.

\end{document}